\documentclass{mn2e}

\usepackage{graphicx}
\usepackage{amsmath}
\usepackage{amssymb}
\usepackage{array}
\usepackage{longtable}
\usepackage{amsfonts}
\usepackage{t1enc}
\usepackage{times}

\newcommand{\Reff}{\mbox{$R_{\rm n}$}}  
\newcommand{\Ie}{\mbox{$I_{\rm n}$}}    
\newcommand{\ec}{\mbox{$e_{\rm c}$}}    
\newcommand{\Sersic}{\mbox{S{\'e}rsic}} 
\newcommand{\etal}{\mbox{et al.}}       
\newcommand{\Ac}{\mbox{$A_{\rm c}$}}    
\newcommand{\Aa}{\mbox{$A_{\rm a}$}}    
\newcommand{\AL}{\mbox{$A_{\rm L}$}}    
\newcommand{\Ab}{\mbox{$A_{\rm b}$}}    
\newcommand{\Bc}{\mbox{$B_{\rm c}$}}    
\newcommand{\Ba}{\mbox{$B_{\rm a}$}}    
\newcommand{\BL}{\mbox{$B_{\rm L}$}}    
\newcommand{\Bb}{\mbox{$B_{\rm b}$}}    
\newcommand{\na}{\mbox{$n_{\rm a}$}}    
\newcommand{\nb}{\mbox{$n_{\rm b}$}}    
\newcommand{\aM}{\mbox{$a_{\rm M}$}}    
\newcommand{\ac}{\mbox{$a_{\rm c}$}}    

\title[The Sérsic profiles along the major and minor axes]
  {The relationship between the Sérsic law profiles measured along the major
  and minor axes of elliptical galaxies\thanks{Partially financed by CNPq}}

\author[F. Ferrari et al. ]{
  F.~Ferrari$^1$\thanks{Programa Institutos do Milênio - MCT/CNPq/PADCT.},
  H.~Dottori$^1$,
  N.~Caon$^2$,
  A.~Nobrega$^{1,3}$,
  D.~B. Pavani$^1$\thanks{CNPq fellow.} \\  \\
  $^1$ Instituto de F\'{\i}sica -- UFRGS, Av. Bento Gon\c{c}alves, 9500,
  Porto Alegre, RS, Brazil. \\
  $^2$ Instituto de Astrofísica de Canarias, Via Lactéa, E-38200 La Laguna,
  Tenerife, Canary Islands, Spain. \\
  $^3$ CETEC -- UNOCHAPECÓ, Av. Senador Att\'{\i}lio Fontana, s/n, Chapec\'o, 
  SC, Brazil.  }
\date{\today}

\begin{document}

\label{firstpage}
\maketitle

\begin{abstract}
  In this paper we discuss the reason why the parameters of the \Sersic\ model
  best-fitting the major axis light profile of elliptical galaxies can differ
  significantly from those derived for the minor axis profile.  We show that
  this discrepancy is a natural consequence of the fact that the isophote
  eccentricity varies with the radius of the isophote and present a mathematical
  transformation that allows the minor axis \Sersic\ model to be calculated from
  the major axis model, provided that the elliptical isophotes are aligned and
  concentric and that their eccentricity can be represented by a well behaved,
  though quite general, function of the radius.  When there is no variation in
  eccentricity only the effective radius changes in the \Sersic\ model, while
  for radial-dependent eccentricity the transformation which allows the minor
  axis \Sersic\ model to be calculated from the major axis model is given by the
  Lerch $\Phi$ transcendental function.  The proposed transformation was tested
  using photometric data for 28 early-type galaxies.
\end{abstract}


\section{Introduction}

It is now recognized that the de Vaucouleurs (1948) $R^{1/4}$ law does not fit
the observed light distribution of elliptical galaxies (e.g.  Schombert 1986).
A much better representation of the light distribution in bright and dwarf
elliptical galaxies and the bulges of spiral galaxies is provided by the
\Sersic\ (1968) law:
\begin{align}
  \label{eq:sersic}
  \log \genfrac{(}{)}{}{}{I(R)}{\Ie}   =   -b_n \left[
  \genfrac{(}{)}{}{}{R}{\Reff}^\frac{1}{n}-1 \right]
\end{align}

\noindent
where \Reff\ is the radius encircling half the total galaxy luminosity and \Ie\ 
is the intensity at \Reff.  The coefficient $b_n$ is a function of $n$, which
can be approximated by the relation $b_n \simeq 2n-0.327$ (Ciotti 1991).

The shape index $n$, which parametrizes the curvature of the \Sersic\ model has
been shown to correlate with the luminosity and size of the galaxy -- brighter
and larger galaxies having larger values of $n$ (Caon, Capaccioli \& D'Onofrio 1993;
subsequently cited as  CCD93) -- and also, notably, with the central velocity dispersion
$\sigma_0$ and the mass of the central supermassive black hole (Graham,
Trujillo \& Caon 2001; Graham \etal\ 2001).

An important source of uncertainty affecting the determination of parameters of
the \Sersic\ model that best describes the light distribution of a galaxy, is on which
axis (major, minor or equivalent) the light profile should be fitted.

CCD93 extensively studied the light profiles of many Virgo cluster E and S0
galaxies by independently fitting \Sersic\ models to their major and minor axes,
finding that in $\sim 40\%$ of the galaxies there were large discrepancies
between the \Sersic\ parameters determined along the major and the minor axes.
Such discrepancies were found not only among S0 galaxies which could be
misclassified as E galaxies but also among genuine elliptical galaxies such as
the E4 galaxy NGC~4621 and E3 galaxy NGC~4406.

Eccentricity gradients imply that both the major and minor axes cannot be, for
example, described by the $R^{1/4}$ model.  The long observed ellipticity
gradients in elliptical galaxies implies that the $R^{1/4}$ model cannot be
universal, but this obvious fact has been largely ignored in the literature.

In this paper we demonstrate that the discrepancy between major and minor axes
\Sersic\ models in elliptical galaxies can be accounted for by radial variations
of the eccentricity of the isophotes.  We also present a mathematical formula
that, coupled with the eccentricity profile, permits transformation of the major
axis \Sersic\ model into the minor axis model, provided that the galaxy has
well-behaved isophotes whose eccentricity varies with radius but which have the
same center and position angle.

In section~\ref{sec:algorithm} we describe the proposed mathematical
transformation, whose applicability and validity is tested by using a sample of
galaxies selected from those studied by CCD93, as described in
section~\ref{sec:sample}.  In section \ref{sec:fitting} we present the fitting
method and in section \ref{sec:results} we analyze and discuss our results.

\section{The link between major and minor axes Sérsic profiles}
\label{sec:algorithm}

A simpler and more convenient representation of the \Sersic\ law is the form 
given in CCD93:
\begin{equation}
  \label{eq:mu}
  \mu{}(R) = A + B \; R^{\frac{1}{n}},
\end{equation}
where, according to equation (\ref{eq:sersic}), 
$A = -2.5\,(b_n+\log \Ie)$, 
$B =  2.5\, b_n/\Reff^{1/n}$.
$R$ may represent the radial variable along the semi major axis $a$, 
the semi minor axis $b$, or the equivalent radius $\sqrt{ab}$. 
The differential of the surface brightness profile can then be written as:
\begin{equation}
  \label{eq:dmu}
  d\mu{}(R) = \frac{B}{n} \; R^{\frac{1}{n} -1} \;  dR.
\end{equation}

\begin{figure}
\begin{center}
\includegraphics[width=0.35\textwidth]{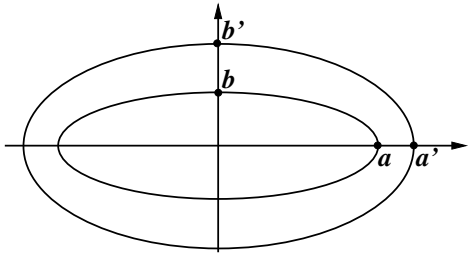}
\caption{Two isophotes with major and minor axes $(a,b)$ and $(a',b')$.}
\label{fig:isophotes}
\end{center}
\end{figure}

Consider two nearby isophotes whose major and minor axes are respectively
$a$ and $b$ for the inner isophote, and $a'$ and $b'$ for the outer one, as
sketched in Fig.~\ref{fig:isophotes}.
The surface brightness gradient along the major axis may be written as:
\begin{equation}
  \label{eq:agrad1}
  \frac{d\mu}{da} = \lim_{\Delta a \to 0} \frac{\mu(a')-\mu(a)}{\Delta a }
\end{equation}
with a similar expression holding true for the minor axis ($b$).  

From the definition of an isophote, we know that $\mu(a)=\mu(b)$ and
$\mu(a')=\mu(b')$, so the numerators in the right hand side of expression
(\ref{eq:agrad1}) and in the equivalent expression for $b$ are equal, while the
denominators $\Delta a$ and $\Delta b$ will differ according to the radial
behavior of the eccentricity\footnote{For analytical simplicity, we use the
  eccentricity $e\equiv b/a$ instead of ellipticity $\varepsilon\equiv 1-e$.}
$e(a)\equiv b/a$. In general we have:
\begin{equation}
  \label{eq:dmu2}
  \frac{d\mu(b)}{db} = \frac{1}{\mathcal{F}(a)} \frac{d\mu(a)}{da}
\end{equation}
where $\mathcal{F}(a)$ will depend on the eccentricity function $e(a)$. 
We discuss the case of constant and variable eccentricity functions in the 
following sections.

\subsection{Constant eccentricity}
\label{sec:e-const}

The simplest case is that of concentric isophotes having constant eccentricity.
If the eccentricity $e \equiv b/a = \ec$ is constant, then we have $b=\ec \,a$
and $db=\ec \,da$, thus:

\begin{equation}
  \label{eq:dmu1}
  \frac{d\mu(b)}{db} = \frac{1}{\ec} \: \frac{d\mu(a)}{da}.
\end{equation}

\begin{figure}
  \centering
  \includegraphics[width=0.45\textwidth]{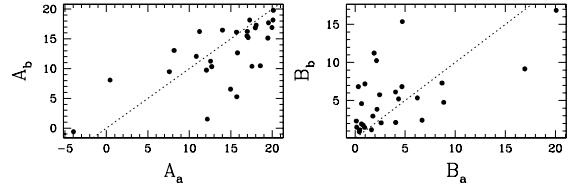}
    \caption{Relationship between the \Sersic\ parameters of equation
      (\ref{eq:mu}). Subscript 'a' refers to the major axis and 
      subscript 'b' to the minor axis.}
  \label{fig:AABB}
\end{figure}

\noindent
By direct integration of equations (\ref{eq:mu}) and (\ref{eq:dmu1}), we see
that in this case the \Sersic\ index $n$ will be the same along the major ($a$)
and the minor ($b$) axes, $\na =\nb$, and that the $B$ coefficients on the major
and minor axes are related by: $\Bb = \Ba/\ec$.  Equation (\ref{eq:dmu1}) shows
that the values of $B$ obtained from the fits along the major and minor axes
should not be considered as independent of each other, as was implicitly
assumed by CCD93 (see Section~\ref{sec:fitting}).  By analyzing the relationship
between $B$ and $b_n$ in equations (\ref{eq:sersic}) and (\ref{eq:mu}), it can
be seen that the effect of \ec\ is to stretch out the relationship between \Ba\ 
and \Bb\ (figure \ref{fig:AABB}).

Theoretically, the integration constants should be equal, i.e. $\Aa\,=\,\Ab$,
since $\mu(a=0)=\mu(b=0)$. However, in real cases (e.g. CCD93) this equality is
broken by a variety of observational uncertainties and practical constraints
(for instance, light profiles are fitted within a surface brightness interval
whose limits in general differ on the major and minor axes).  As a consequence,
different values for \Aa\ and \Ab\ are obtained when the fitted profile is
extrapolated to $R = 0$.

The \Sersic\ model along the minor axis is related to the \Sersic\ model
along the major axis by the equation:
\begin{equation}
  \label{eq:ec-mub}
  \mu(b) = \Aa + \frac{\Ba}{\ec} \; b^{1/\na}
\end{equation}
where \na\ is the major axis \Sersic\ index.

\subsection{Variable eccentricity}
\label{sec:e-var}

In most galaxies, eccentricity is neither constant, nor is it a simple function
of the radius. Indeed, no general rules seem to govern the radial variation of
$e$, and it is not clear what the physical significance of this variation is
(Binney \& Merriefield 1998).  In cD galaxies, $e$ generally decreases from the
center outwards, while in other galaxies $e(R)$ may increase, and sometimes it
is found to vary non-monotonically with the radius.

Now, if the eccentricity is a differentiable function $e=e(a)$, then 
$db=e(a)\, da + a\, de$ or, equivalently, 
\begin{equation}
db
=
\left[ e(a) + a \, \frac{de}{da} \right] \, da
\equiv
\mathcal{F}(a) \,  da.
\label{eq:dbda}
\end{equation}

\noindent
In this case, the minor axis profile may have a shape very different from that 
of the major axis, depending on the form of function $e(a)$.  
We have integrated equation~(\ref{eq:dmu2}) for a general case in which 
$e(a)$ can be expressed as a function of the form
\begin{equation}
  \label{eq:epsilon}
  e(a) = e_0 + (e_1 - e_0)
  \genfrac{(}{)}{}{}{a}{\aM}^{l},
\end{equation}

\noindent
where \aM\ is the scale length where the eccentricity equals $e_1$. 
Depending on $l$, $e_0$ and $e_1$, equation (\ref{eq:epsilon}) may describe 
radial increasing ($e_0\,<\,e_1$) or decreasing ($e_0\,>\,e_1$) eccentricities,
with different slopes. 
From equations (\ref{eq:dbda}) and (\ref{eq:epsilon}) we can derive:
\begin{equation}
  \label{eq:dbepsilon}
  \frac{db}{da} = e_0 + (1+l)(e_1-e_0)
  \genfrac{(}{)}{}{}{a}{\aM}^l ,
\end{equation}
from which it follows
\begin{equation}
  \label{eq:dmudbepsilon}
  \frac{d\mu{}}{db} =
  \left[ e_0 + (1+l) (e_1-e_0) \genfrac{(}{)}{}{}{a}{\aM}^l \right]^{-1}
  \frac{d\mu{}}{da}.
\end{equation}
By comparing equations (\ref{eq:dmu2}) and (\ref{eq:dmudbepsilon}) we obtain
$\mathcal{F}(a)$,
\begin{align}
  \label{eq:Fa}
  \mathcal{F}(a)\,=\, e_0 + (1+l) (e_1-e_0)
  \genfrac{(}{)}{}{}{a}{\aM}^l
\end{align}
We can integrate equation (\ref{eq:dmudbepsilon}) in terms of the 
transcendental function Lerch $\Phi$ (Gradshteyn \& Ryzhik 2000, see appendix 
A), obtaining:
\begin{equation}
  \label{eq:muphi}
  \mu_{\rm L}(b) =  \Aa + \frac{\Ba}{e_0 \; \na\; l} \;\;  
  a^{1/\na} \;\;
  \Phi \left( 1 - \frac{ \mathcal{F}(a)}{e_0}  \; ;
  \;  1 \; ;  \;  \frac{1}{{\na}\;l} \right)
\end{equation}
The variable $b$ does not appear explicitly on the right-hand side of this
equation; in order to compute $\mu$ at a given distance $b'$ on the minor axis,
we should set the variable $a$ on the right-hand side to that value of $a'$ for
which $b' = e(a')\,a'$.

Equation (\ref{eq:muphi}) shows how the major axis \Sersic\ law is modulated by
the Lerch $\Phi$ function to give the minor axis light profile. By comparing it
with the \Sersic\ law for the minor axis: \mbox{$\mu{}(b) = \Ab + \Bb \;
  b^{1/\nb}\;$} (equation~\ref{eq:mu}), we can write:
\begin{eqnarray}
\label{eq:bteo}
\Ab  & \;  \Longleftrightarrow \; &  \Aa \nonumber \\
\Bb  & \;  \Longleftrightarrow \; &  \frac{\Ba}{e_0 \; \na \; l} \nonumber \\
b^{{1}/{\nb}}                &
\;  \Longleftrightarrow \;     &
a^{1/\na} \; \Phi \left(1- \frac{\mathcal{F}(a)}{e_0 } \; ;
\;  1 \; ;  \; \frac{1}{{\na} \: l} \right) 
\end{eqnarray}

\subsection{The equivalent-axis Sérsic profile}
\label{sec:R}

The \Sersic\ law can also be expressed as a function of the equivalent radius,
defined as $R_{\rm eq} = \sqrt{a b}$. In the case of constant eccentricity, 
$e(a) = \ec = {\rm const.}$, equation (\ref{eq:dmu1}) can be written as:
\begin{align}
  \label{eq:dmudbR}
  \frac{d\mu{}}{d R_{\rm eq} } = \frac{1}{\sqrt{\ec}} \frac{d\mu{}(a)}{da},
\end{align}
while, for variable eccentricity, equation (\ref{eq:dmudbepsilon}) can be
expressed as:
\begin{equation}
  \label{eq:dmudbeR}
  \frac{d\mu{}}{d R_{\rm eq} } =
  \frac{2\sqrt{e(a)}}{e(a) + \mathcal{F}(a)} \:\:
  \frac{d\mu{}}{da},
\end{equation}
where $e(a)$ is given by equation (\ref{eq:epsilon}) and $\mathcal{F}(a)$ by
equation (\ref{eq:Fa}). We were not able to integrate equation
(\ref{eq:dmudbeR}).

\section{Data set used}
\label{sec:sample}

We applied the algorithm developed in the previous section to 28 galaxies 
selected from those studied by CCD93. Surface brightness and ellipticity 
profiles for these objects were published by Caon, Capaccioli \& Rampazzo
(1990), and Caon, Capaccioli \& D'Onofrio (1994).
The sample we use covered a wide interval of absolute magnitudes 
($-22.43<M_B<-17.29$) and included at least one object for each morphological 
type (E0 to E7, dS0 and S0). 

The correspondence between the \Sersic\ model index $n$ for the major ($a$) and
minor ($b$) axis also varied: $\na > \nb$ for 8 galaxies; $\na < \nb$ for 17,
and $\na \simeq \nb$ for 3.  The eccentricity (Figures \ref{fig:elperfA} and
\ref{fig:elperfB}) increased with radius for 12 objects, decreased for another
12 and remained approximately constant for 4.  The central parts of the light
profiles, affected by seeing convolution, were excluded when fitting our
eccentricity model (equation \ref{eq:epsilon}) to the observed profiles.

\subsection{Errors}
\label{sec:errors}

The photometric uncertainties on the CCD93 $B$-band surface brightness 
measurements were estimated by Caon \etal\ (1990), and are shown in Figure 3 
of their paper. 
They can be approximated by the power-law function:
\begin{equation}
  \label{eq:error}
  \delta\mu = \alpha \: \mu^{\beta}
\end{equation}
where $\delta\mu$ is the error, $\mu$ the surface brightness in  magnitudes, 
$\alpha\simeq 3.25\,10^{-15}$ and $\beta\simeq 9.7$.

The error in the eccentricity can be estimated by approximating the 
differentials in equation (\ref{eq:dmu}) by small variations, i.e., 
$d\mu\approx \delta\mu$ and $dR\approx \delta R$, thus obtaining 
$\delta\mu{} = (B/n) \; R^{\frac{1}{n} -1} \; \delta R$. 
Rearranging the terms with the help of equation (\ref{eq:error}) we can write 
the fractional error $\delta R/R$ as:
\begin{equation}
  \label{eq:deltaR}
  \frac{\delta R}{R} = \frac{n \: \alpha \: \mu^\beta }{B\: R^{\frac{1}{n}} }
  =
  \frac{n \: \alpha \: (A+B\:R^{\frac{1}{n}})^\beta }{B\: R^{\frac{1}{n}} }.
\end{equation}
Here $R$ may be the $a$ or $b$ variable and the coefficients $A, B, n$ may 
refer to the major or  minor axis accordingly.  
Since the eccentricity is calculated as the quotient $b/a$, the fractional 
uncertainties add to give:
\begin{equation}
  \label{eq:eccerr}
  \frac{\delta e}{e} \approx \frac{\delta a}{a} + \frac{\delta b}{b} .
\end{equation}
For example, in the outer parts ($a=296'', \; b=180''$) of NGC 4473 we have
$\delta\mu(a)\simeq 0.31 \;{\rm mag/arcsec^2}$, 
$\delta\mu(b)\simeq 0.43 \;{\rm mag/arcsec^2}$, 
$\delta a/a\simeq 0.08$ and $\delta b/b\simeq 0.15$, which
yields $\delta e/e\simeq 0.23$.  
For NGC 4406 ($a=510'',\; b=330''$), 
$\delta\mu(a) = 0.17 \;{\rm mag/arcsec^2}$, 
$\delta\mu(b) = 0.28\;{\rm mag/arcsec^2}$, 
$\delta a/a \simeq 0.06$ and $\delta b/b \simeq 0.11$,
thus $\delta e /e \simeq 0.17$.

\section{Fitting method}
\label{sec:fitting}

For each of the 28 galaxies of the sample, a Levenberg-Marquardt algorithm was
used to fit the minor axis surface brightness profile using the transformed
major axis Sérsic law. The data for the major and minor axes light profiles are
those analyzed by CCD93.

The fit was done for both the approximation of constant eccentricity, and 
for the more general case of variable eccentricity. We use the following 
notation: 
\begin{equation}
  \label{eq:muc}
  \mu_{\rm c} = \Ac + \frac{\Bc}{\ec} \: a^{1/\na}
\end{equation}
and
\begin{equation}
  \label{eq:muL}
  \mu_{\rm L} = \AL + \frac{B_L}{e_0\, \na \, l} \; a^{1/\na} \;
  \Phi \left(1- \frac{ \mathcal{F}(a)}{e_0}   \; ;
   \;  1 \; ;  \; \frac{1}{{\na} \: l} \right).
\end{equation}
Equation (\ref{eq:muc}) is for constant eccentricity and equation (\ref{eq:muL})
is for variable eccentricity.

We decided to leave the parameters $A$ and $B$ completely free. The parameters
\na\ is the major axis \Sersic\ index measured by CCD93, while the parameters
$e_0$ and $l$ and the function $\mathcal{F}(a)$ are set by our fit to the
eccentricity profiles. We noticed that, ideally, the values we obtained for \Ac\ 
and \AL\ should  equal  \Aa while the values for \Bc\ and \BL\ should be
equal to \Ba\ (where \Aa\ and \Ba\ are the values measured by CCD93.)  Thus, the
validity of our results, and hence of our proposed method, is determined by how
close the above parameters are to their expected values.

The parameters obtained by fitting equations (\ref{eq:muc}) and (\ref{eq:muL}) 
to the CCD93 minor axis profiles are listed in Table~\ref{tab:results}, where
for comparison we include the parameters found by CCD93.

In appendix \ref{sec:profiles} we present the results of 
Table~\ref{tab:results} in graphical format; these figures also show the major 
axis profile. The bottom panel shows the residuals between the CCD93 data 
and our best-fitting model.

In appendix \ref{sec:elperf} we present the fits to the eccentricity profiles
(Figures \ref{fig:elperfA} and \ref{fig:elperfB}) derived from CCD93 data, the
solid lines showing the least-squares fit of the function given by equation
(\ref{eq:epsilon}) to the data points.  For some galaxies, we could not use the
parameters obtained by this fit and had to determine them interactively. In
fact, the Lerch $\Phi$ critical radius \ac\ (Appendix \ref{sec:lerch}) must be
larger than the largest observed radius for the Lerch $\Phi$ function to
converge in the radial interval covered by CCD93 observations.  The eccentricity
profile parameters are shown in Table~\ref{tab:elperf}.

\begin{table}
\begin{center}
\begin{tabular}{l r >{$}r<{$}  >{$}r<{$}  >{$}r<{$}  >{$}r<{$}   >{$}r<{$} } 
\hline
Galaxy   & Type  & e_0  & e_1  & \ec\  & l    & \aM\  \\\hline
NGC 4168 & E2    & 0.88 & 0.78 & 0.83 & 1.50 & 120  \\
NGC 4261 & E2    & 0.84 & 0.75 & 0.78 & 0.75 & 250  \\
NGC 4339 & S0(0) & 0.95 & 0.86 & 0.93 & 2.00 & 120  \\
NGC 4360 & E2    & 0.90 & 0.75 & 0.81 & 0.80 & 120  \\
NGC 4365 & E3    & 0.75 & 0.66 & 0.74 & 3.50 & 300  \\
NGC 4374 & E1    & 0.70 & 0.96 & 0.93 & 0.15 & 380  \\
NGC 4387 & E5    & 0.60 & 0.76 & 0.80 & 1.22 & 110  \\
NGC 4406 & E3    & 0.88 & 0.57 & 0.65 & 0.35 & 700  \\
NGC 4415 & dE1,N & 0.90 & 0.86 & 0.89 & 1.00 &  80  \\
NGC 4431 & dS0,N & 0.53 & 0.75 & 0.65 & 1.35 &  72  \\
NGC 4434 & E0    & 0.96 & 0.82 & 0.95 & 2.50 &  84  \\
NGC 4436 & dS0,N & 0.47 & 0.60 & 0.70 & 2.50 & 110  \\
NGC 4458 & E1    & 0.84 & 0.98 & 0.90 & 0.51 &  90  \\
NGC 4472 & E2    & 1.00 & 0.75 & 0.80 & 0.16 & 715  \\
NGC 4473 & E5    & 0.45 & 0.69 & 0.60 & 0.55 & 330  \\
NGC 4476 & S0(5) & 0.58 & 0.91 & 0.85 & 0.75 & 154  \\
NGC 4478 & E2    & 0.82 & 0.97 & 0.88 & 3.00 &  77  \\
NGC 4486 & E0    & 1.00 & 0.60 & 0.85 & 0.65 & 550  \\
NGC 4550 & S0(7) & 0.39 & 0.22 & 0.30 & 0.78 & 154  \\
NGC 4551 & E2    & 0.68 & 0.82 & 0.75 & 1.00 &  85  \\
NGC 4552 & S0(0) & 1.00 & 0.81 & 0.88 & 0.43 & 300  \\
NGC 4564 & E6    & 0.44 & 0.61 & 0.60 & 1.00 & 190  \\
NGC 4600 & S0(6) & 0.62 & 0.85 & 0.80 & 1.00 &  77  \\
NGC 4621 & E4    & 0.65 & 0.95 & 0.90 & 1.00 & 360  \\
NGC 4623 & E7    & 0.90 & 0.22 & 0.41 & 0.15 & 110  \\
NGC 4636 & E1    & 1.00 & 0.62 & 0.72 & 0.39 & 400  \\
NGC 4649 & S0(2) & 0.77 & 0.83 & 0.82 & 0.60 & 640  \\
NGC 4660 & E3    & 0.55 & 0.86 & 0.82 & 0.70 & 130  \\
\hline
\end{tabular}
\caption{Galaxy name, type and the eccentricity profile parameters.
$e_0$ is the eccentricity at $a=0$, $e_1$ the eccentricity at $a=\aM$, \ec\ is 
the value for the case of constant eccentricity, and $l$ is the exponent.}
\label{tab:elperf}
\end{center}
\end{table}

\section{The results}
\label{sec:results}

The analysis of the results shown in Table~\ref{tab:results} reveals an 
overall good agreement between the computed and the expected values.

For 14 of the galaxies, both \Ac\ (the zero point in the constant eccentricity
model) and \AL\ (the zero point in the variable eccentricity model) differ by
less than 0.5 mag from the best fit \Aa\ values determined by CCD93. For further
8 galaxies the difference for both coefficients is less than 1 mag.  The most
discrepant galaxies are NGC~4406, NGC~4374 and NGC~4552 for which $|\AL-Aa| >
1.5$ mag.

As for scale lengths (the $B$ parameters in Table \ref{tab:results}), 15
galaxies have \Bc\ and \BL\ values which both differ by less than 20\% from
\Ba, while for 8 galaxies the difference is less than 30\%, the most discrepant
object being NGC~4564 for which $|\Bc-\Ba|\,/\,\Ba = 0.38$.

\begin{figure}
  \centering
    \includegraphics[width=0.45\textwidth]{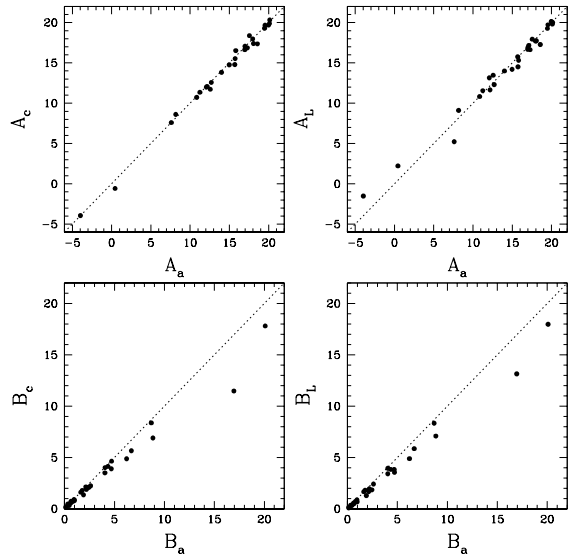}
    \caption{Relationship between the major axis parameters from
     CCD93 (\Aa; \Ba) and the parameters \Ac, \Bc, \AL\ and \BL\ derived 
     in this paper. The scatter observed in Figure \ref{fig:AABB} is here 
     greatly reduced.}
    \label{fig:AABB2}
\end{figure}

Figure~\ref{fig:AABB2} shows how the minor axis \Sersic\ parameters derived
using our method correlates well with the major axis parameters, this new
correlation being a remarkable improvement over that shown in
Figure~\ref{fig:AABB}.  The fact that the values of \Ac, \AL, \Bc, \BL\ are
close to their expected values (\Aa\ and \Ba) indicates that our transformed
major axis \Sersic\ models can fit the minor axis light profiles quite well.

These results support our proposal that the differences in the \Sersic\ model of
the major and minor axes can be accounted for by radial variations of the
isophotes eccentricity, indeed our model seems to be able to provide a valid
mathematical description of the links between major and minor axes light
profiles and the eccentricity profile.

\begin{table*}
  \begin{center}
    \begin{tabular}{l >{$}r<{$}  >{$}r<{$}  >{$}r<{$}  >{$}r<{$}   >{$}r<{$} 
                      >{$}r<{$}  >{$}r<{$}  >{$}r<{$}  >{$}r<{$}   >{$}r<{$} 
                      >{$}r<{$}  >{$}r<{$} >{$}r<{$}} \hline
     Galaxy    & \Aa\  & \Ab\  & \Ac\   & \AL\   & &  \Ba\   &
                 \Bb\  & \Bc\  & \BL\ & & {\rm RMS}_{\rm b} & 
                 {\rm RMS}_{\rm c} & {\rm RMS}_{\rm L} 
                 \\\hline
NGC 4168  & 11.24 & 16.22 & 11.36 & 11.55 &&  6.689 &  2.406 &  5.650 &  5.862 && 0.205 & 0.722 & 0.613 \\
NGC 4261  &  8.17 & 13.05 &  8.61 &  9.11 &&  8.852 &  4.760 &  6.900 &  7.085 && 0.692 & 0.892 & 0.666 \\
NGC 4339  & 16.96 & 15.49 & 16.64 & 16.68 &&  1.772 &  2.950 &  1.790 &  1.806 && 0.461 & 0.759 & 0.825 \\
NGC 4360  & 17.10 & 15.21 & 16.88 & 17.15 &&  2.190 &  3.850 &  1.894 &  1.950 && 0.459 & 0.827 & 1.156 \\
NGC 4365  & 10.84 & 12.05 & 10.73 & 10.83 &&  6.215 &  5.349 &  4.876 &  4.887 && 0.473 & 0.626 & 0.460 \\
NGC 4374  &  7.60 &  9.48 &  7.60 &  5.23 &&  8.682 &  7.299 &  8.373 &  8.338 && 0.339 & 0.315 & 0.458 \\
NGC 4387  & 17.55 & 10.39 & 18.38 & 17.94 &&  0.983 &  7.191 &  0.795 &  0.682 && 0.838 & 2.527 & 1.695 \\
NGC 4406  &  0.44 &  8.08 & -0.57 &  2.23 && 16.950 &  9.159 & 11.477 & 13.142 && 0.394 & 0.805 & 0.334 \\
NGC 4415  & 19.55 & 17.69 & 19.68 & 19.72 &&  0.640 &  1.921 &  0.576 &  0.568 && 1.407 & 0.669 & 0.754 \\
NGC 4431  & 20.14 & 19.79 & 20.32 & 20.02 &&  0.418 &  0.894 &  0.345 &  0.337 && 0.459 & 0.923 & 0.399 \\
NGC 4434  & 14.00 & 16.46 & 13.83 & 14.00 &&  4.053 &  2.111 &  4.011 &  3.945 && 0.996 & 0.639 & 0.455 \\
NGC 4436  & 19.48 & 15.13 & 19.30 & 19.28 &&  0.643 &  4.595 &  0.732 &  0.494 && 0.633 & 0.554 & 0.530 \\
NGC 4458  & 17.28 & 18.17 & 16.86 & 16.64 &&  1.635 &  1.168 &  1.614 &  1.661 && 3.313 & 1.399 & 1.130 \\
NGC 4472  & 12.07 &  9.75 & 11.99 & 13.15 &&  4.680 &  6.814 &  3.905 &  3.833 && 0.258 & 0.435 & 0.909 \\
NGC 4473  & 15.73 &  5.27 & 15.53 & 14.52 &&  1.897 & 11.238 &  1.360 &  1.287 && 0.330 & 1.872 & 0.957 \\
NGC 4476  & 15.81 & 12.64 & 16.52 & 15.32 &&  2.446 &  5.759 &  2.089 &  1.842 && 0.885 & 1.440 & 0.779 \\
NGC 4478  & 16.99 & 16.25 & 17.07 & 16.92 &&  0.954 &  1.474 &  0.904 &  0.883 && 0.883 & 0.481 & 0.397 \\
NGC 4486  & 12.57 & 11.25 & 11.73 & 13.47 &&  4.346 &  5.215 &  4.129 &  3.836 && 0.467 & 0.409 & 0.983 \\
NGC 4550  & 18.07 & 17.30 & 17.41 & 17.75 &&  0.468 &  1.149 &  0.321 &  0.359 && 0.758 & 0.584 & 0.923 \\
NGC 4551  & 17.97 & 16.87 & 17.97 & 17.71 &&  0.816 &  1.754 &  0.711 &  0.719 && 1.079 & 0.840 & 0.620 \\
NGC 4552  & -3.97 & -0.61 & -3.94 & -1.52 && 20.087 & 16.850 & 17.816 & 17.982 && 1.205 & 1.241 & 1.040 \\
NGC 4564  & 18.57 & 10.48 & 17.36 & 17.28 &&  0.329 &  6.819 &  0.454 &  0.349 && 0.298 & 1.796 & 1.667 \\
NGC 4600  & 20.10 & 18.16 & 19.89 & 19.83 &&  0.163 &  1.484 &  0.206 &  0.175 && 0.550 & 0.567 & 0.442 \\
NGC 4621  & 12.17 &  1.52 & 12.07 & 11.66 &&  4.714 & 15.363 &  4.641 &  3.556 && 0.590 & 0.590 & 0.293 \\
NGC 4623  & 19.98 & 16.89 & 19.68 & 20.15 &&  0.130 &  2.302 &  0.130 &  0.096 && 0.104 & 2.588 & 3.749 \\
NGC 4636  & 15.69 & 16.13 & 14.80 & 15.75 &&  2.608 &  2.069 &  2.246 &  2.407 && 1.054 & 1.062 & 1.203 \\
NGC 4649  & 12.70 & 10.34 & 12.58 & 12.31 &&  4.038 &  6.122 &  3.499 &  3.416 && 0.797 & 0.725 & 0.595 \\
NGC 4660  & 14.98 &  6.55 & 14.76 & 14.20 &&  2.140 & 10.251 &  2.133 &  1.671 && 0.711 & 0.976 & 0.476 \\
\hline
    \end{tabular}
    \caption{Best-fit \Sersic\ parameters (following the notation in equation
      (\ref{eq:mu})): zero point $A$ and scale length $B$.  \Aa, \Ab, \Ba\ and
      \Bb\ are the parameters measured by CCD93 on major (subscript 'a') and
      minor (subscript 'b') axes.  \Ac, \AL, \Bc\ and \BL\ are the parameters
      computed by us for constant (subscript 'c') and variable (subscript
      'L') eccentricity.  The root mean square (RMS) residuals of the fits are
      shown in the last three columns.}
    \label{tab:results}
  \end{center}
\end{table*}

There is increasing interest in using the $R^{\frac{1}{n}}$ law to address some
issues related to the fundamental plane (FP) of elliptical galaxies (Ciotti,
Lanzoni \& Renzini 1996; Graham \& Colless 1997; Ciotti \& Lanzoni 1997), thus
an extension of the work presented in our current paper would be to investigate
how fitting the \Sersic\ model on different axes may affect the distribution of
galaxies on the fundamental plane.  This is because two galaxies with the same
major axis light profile, but different eccentricity profiles, can give
different values for the index $n$ when the \Sersic\ model is fitted to their
equivalent axis profile. This is because $R_{\rm eq}=\sqrt{ab}=a\,\sqrt{e(a)}$,
which may account for some of the scatter observed in the fundamental plane.  A
full study of this topic is, however, outside the scope of the present paper.

\pagebreak

\appendix{}

\section{Eccentricity profiles}
\label{sec:elperf}

\begin{figure}
  \begin{center}
   \includegraphics[width=.45\textwidth]{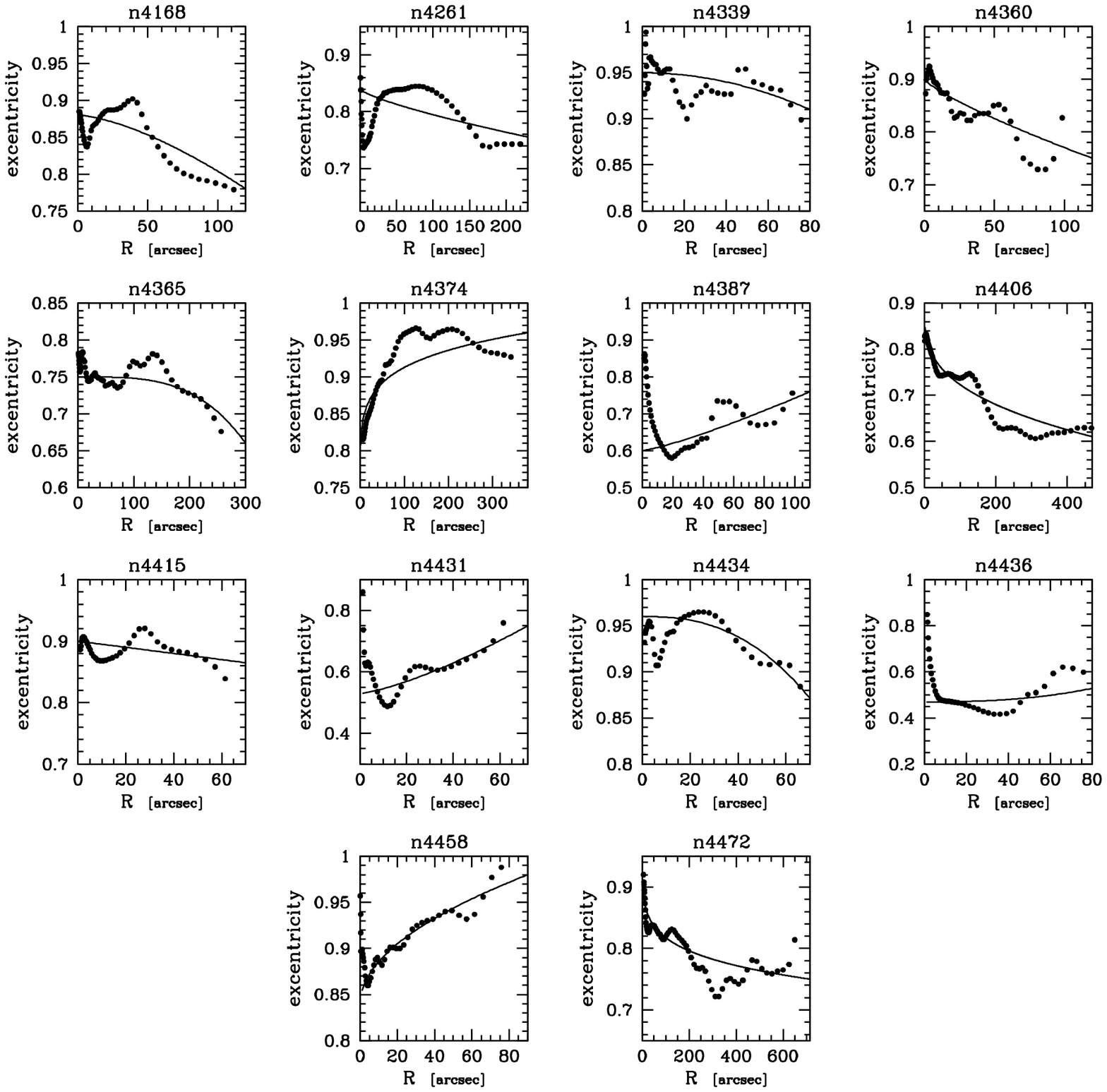}
   \caption{Eccentricity profiles. The dotted line shows the observed 
   eccentricity from CCD93 data; the solid line is the least-square fit
   of formula (\ref{eq:epsilon}) to the data.}
   \label{fig:elperfA}
  \end{center}
\end{figure}

\begin{figure}
  \begin{center}
    \includegraphics[width=.45\textwidth]{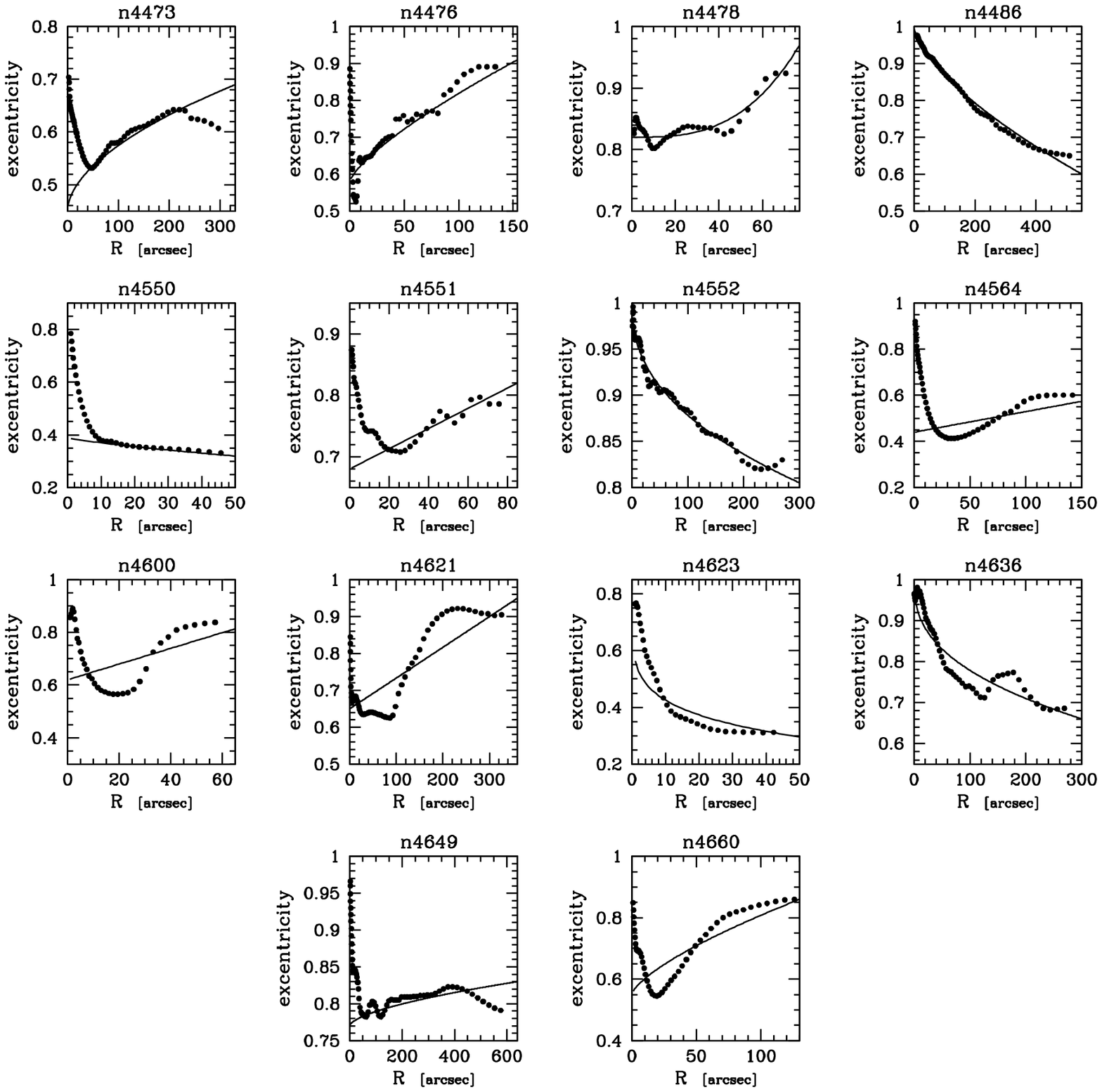}
    \caption{Eccentricity profiles. Continued.}
    \label{fig:elperfB}
  \end{center}
\end{figure}

\pagebreak

\section{Brightness profiles}
\label{sec:profiles}

\begin{figure}
  \begin{center}
    \label{fig:mugalA}
    \includegraphics[width=.5\textwidth]{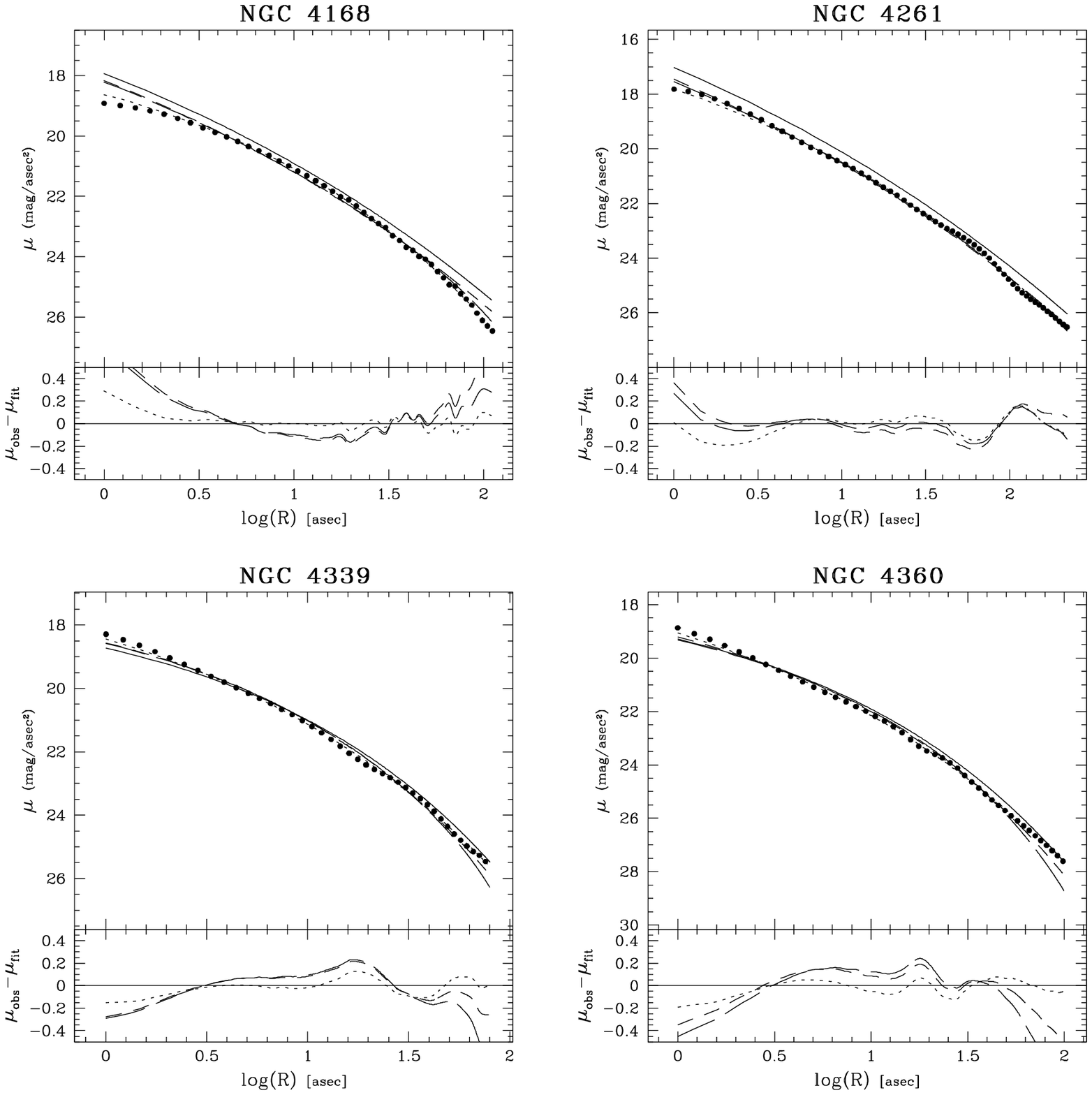}
    \caption{Surface brightness profiles. Solid and  dotted lines 
    represent the CCD93 \Sersic\ fits to the galaxies major and minor axes 
    profiles respectively; the short and long dashed lines represent our 
    transformation of the major axis S\'ersic law by constant and variable 
    eccentricity, respectively. The bottom panel shows the residuals between 
    the CCD93 data and the best-fit models, using the same line styles 
    as described above.}
  \end{center}
\end{figure}

\begin{figure}
  \begin{center}
    \label{fig:mugalB}
    \includegraphics[width=.5\textwidth]{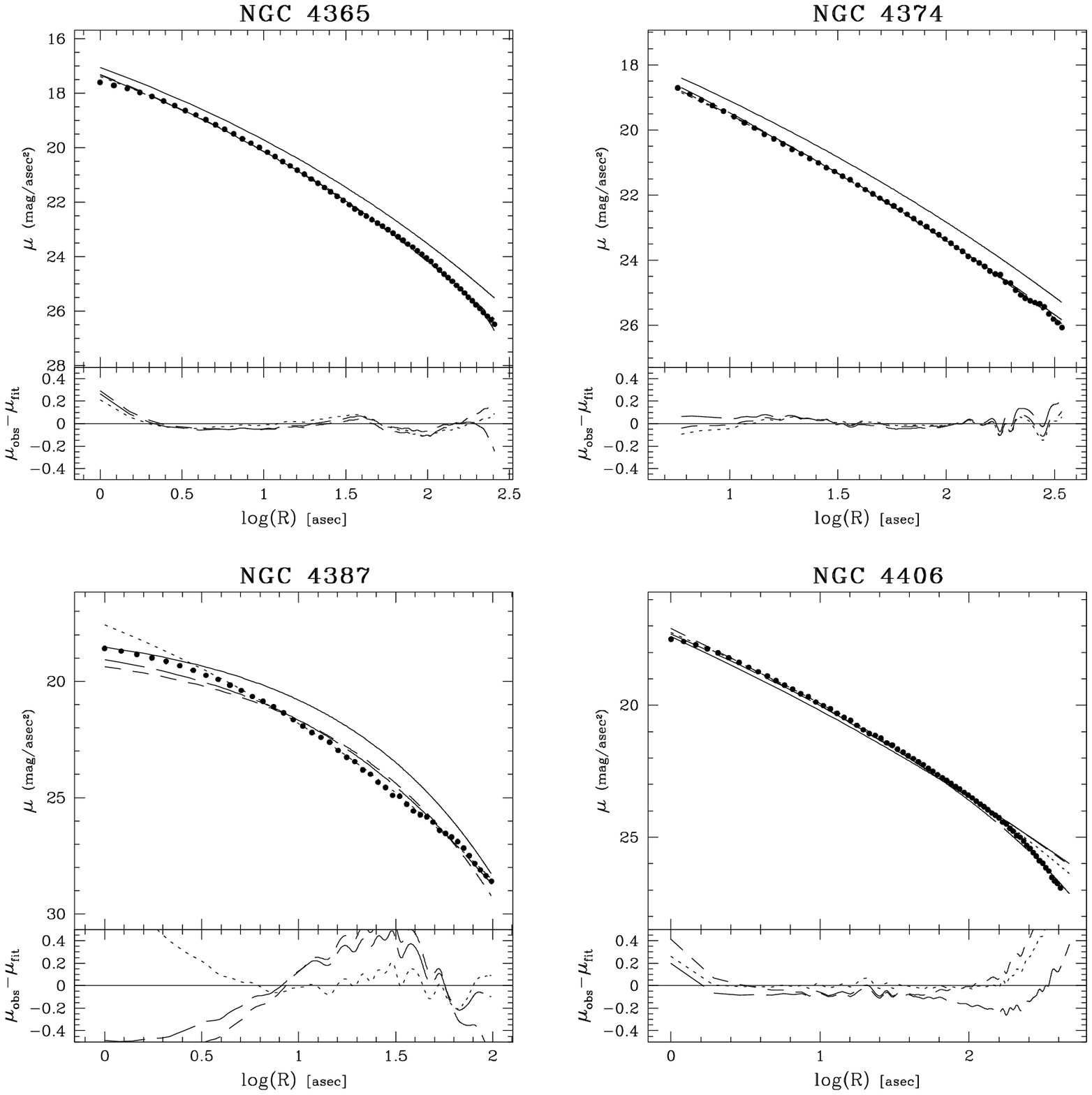}
    \caption{Surface brightness profiles. Continued.}
  \end{center}
\end{figure}

\begin{figure}
  \begin{center}
    \label{fig:mugalC}
    \includegraphics[width=0.5\textwidth]{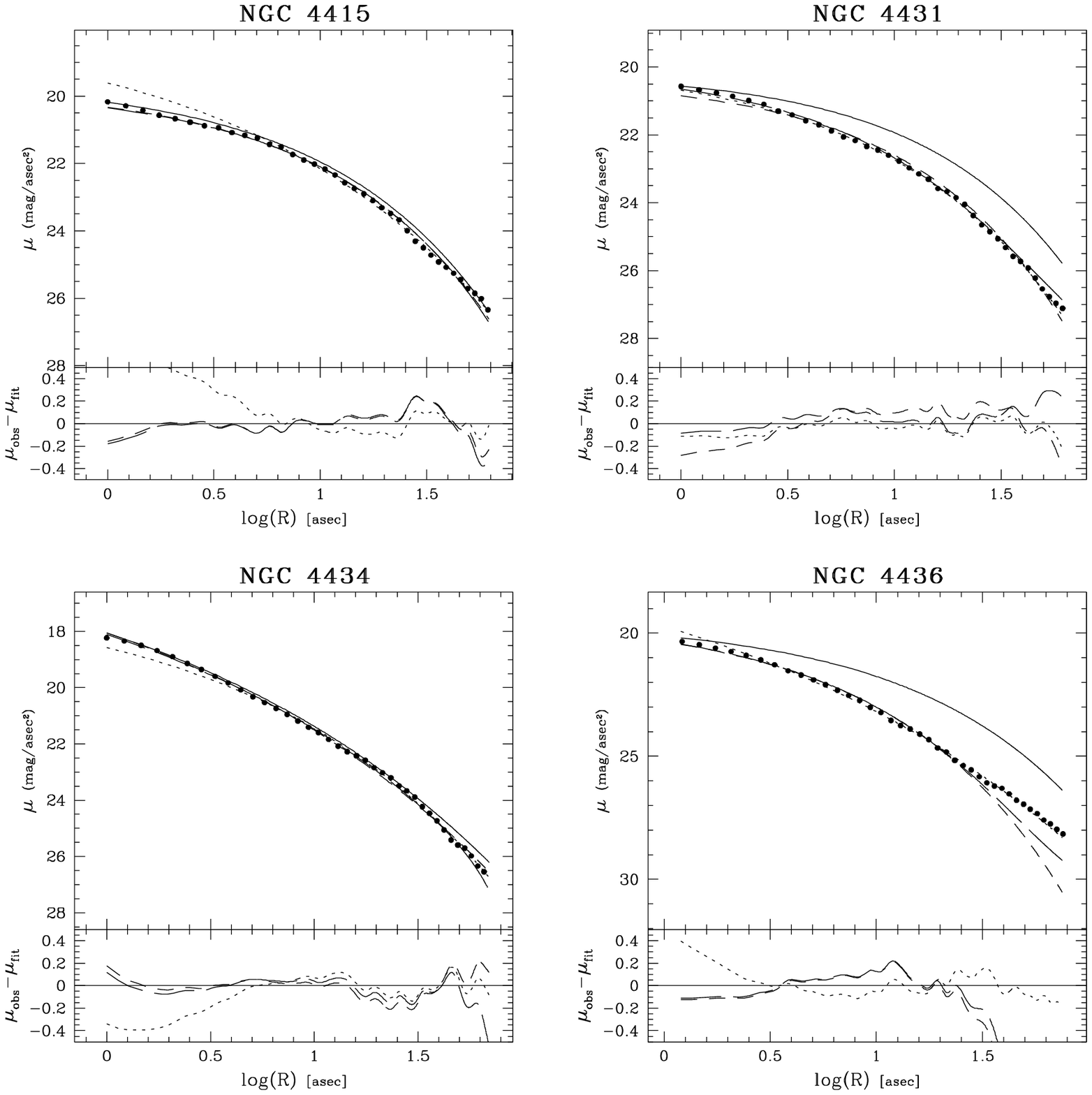}
    \caption{Surface brightness profiles. Continued.}
  \end{center}
\end{figure}

\begin{figure}
  \begin{center}
    \label{fig:mugalD}
    \includegraphics[width=0.5\textwidth]{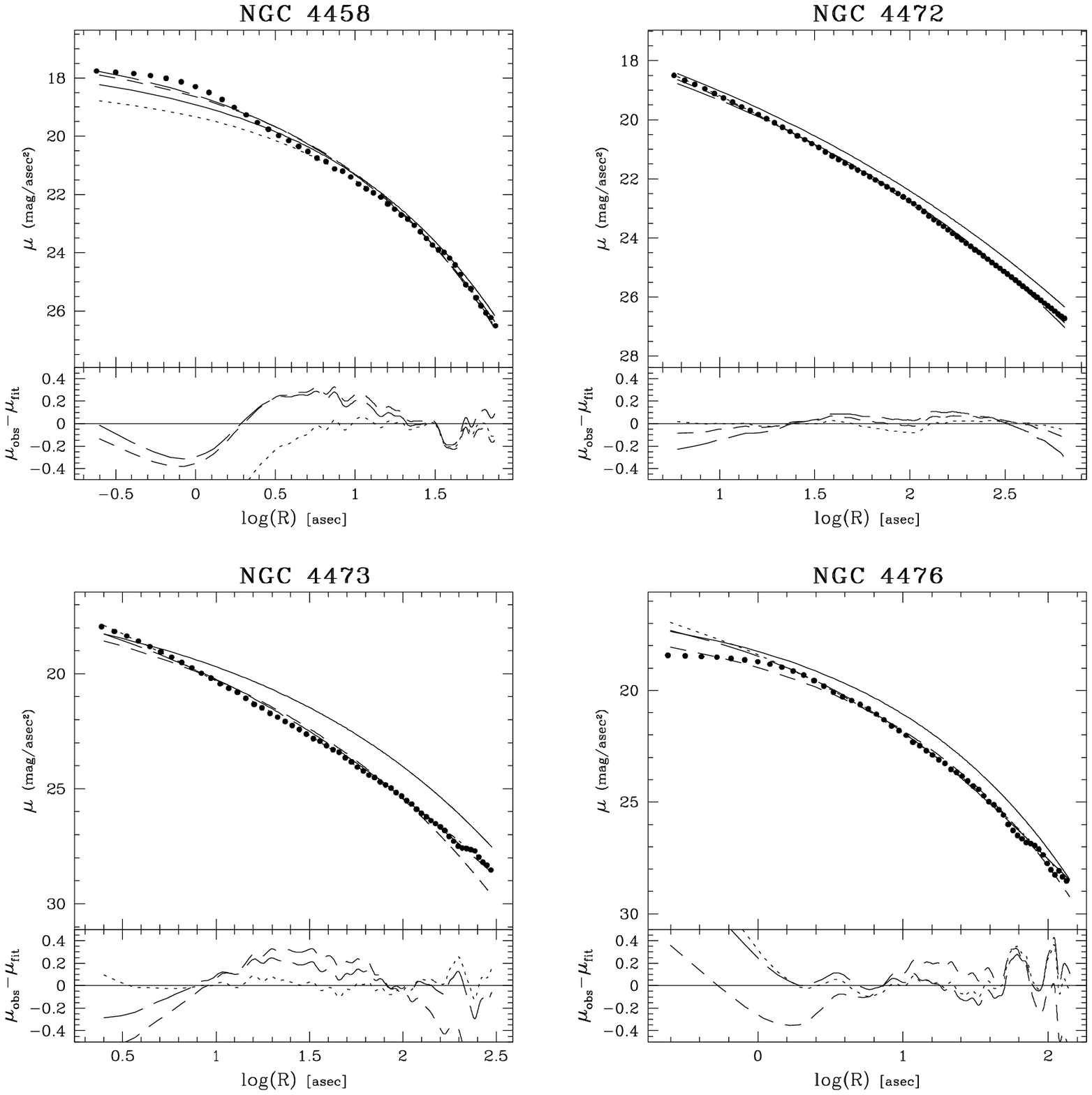}
    \caption{Surface brightness profiles. Continued.}
  \end{center}
\end{figure}

\begin{figure}
  \begin{center}
    \label{fig:mugalE}
    \includegraphics[width=0.5\textwidth]{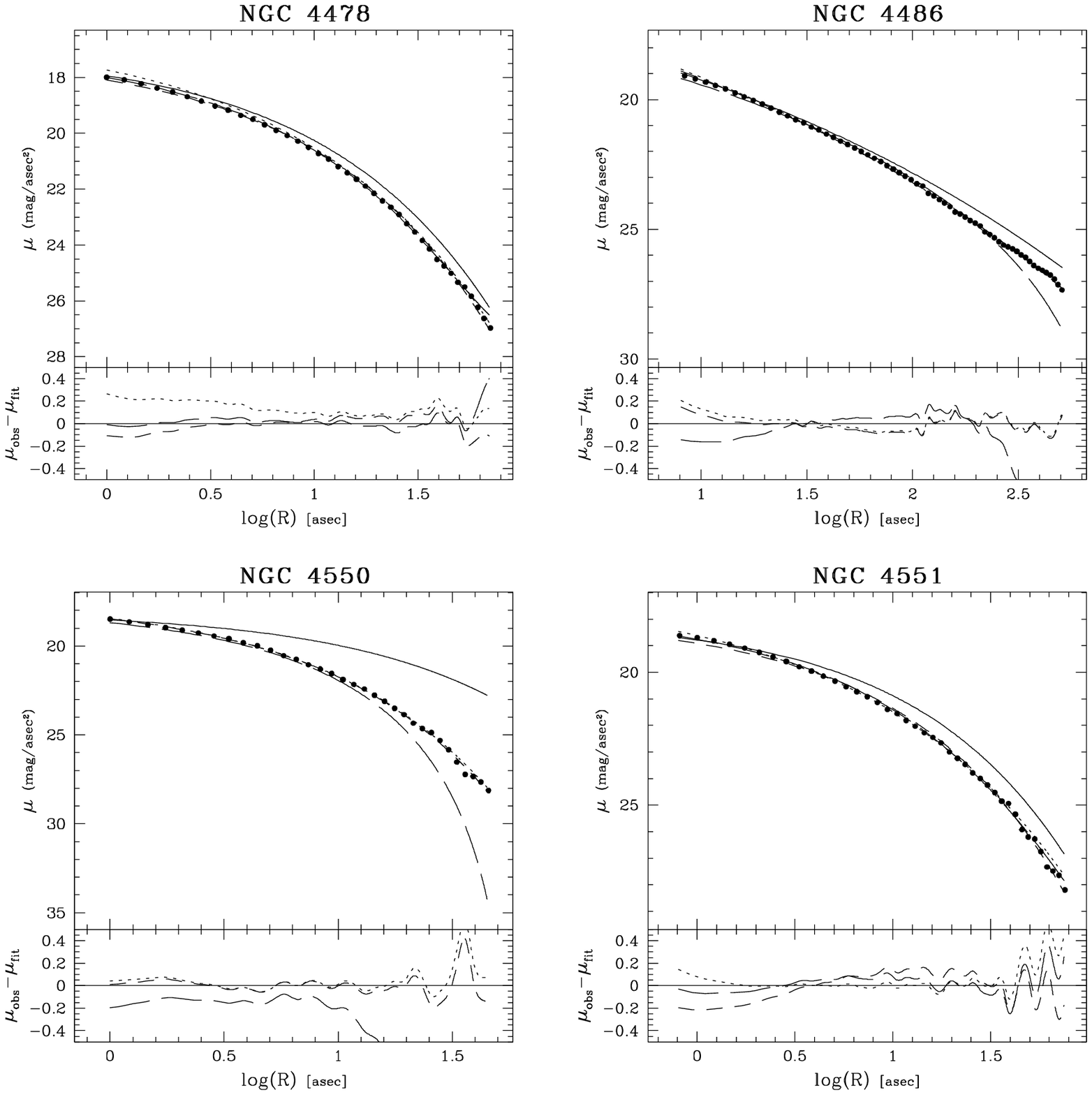}
    \caption{Surface brightness profiles. Continued.}
  \end{center}
\end{figure}

\begin{figure}
  \begin{center}
    \label{fig:mugalF}
    \includegraphics[width=0.5\textwidth]{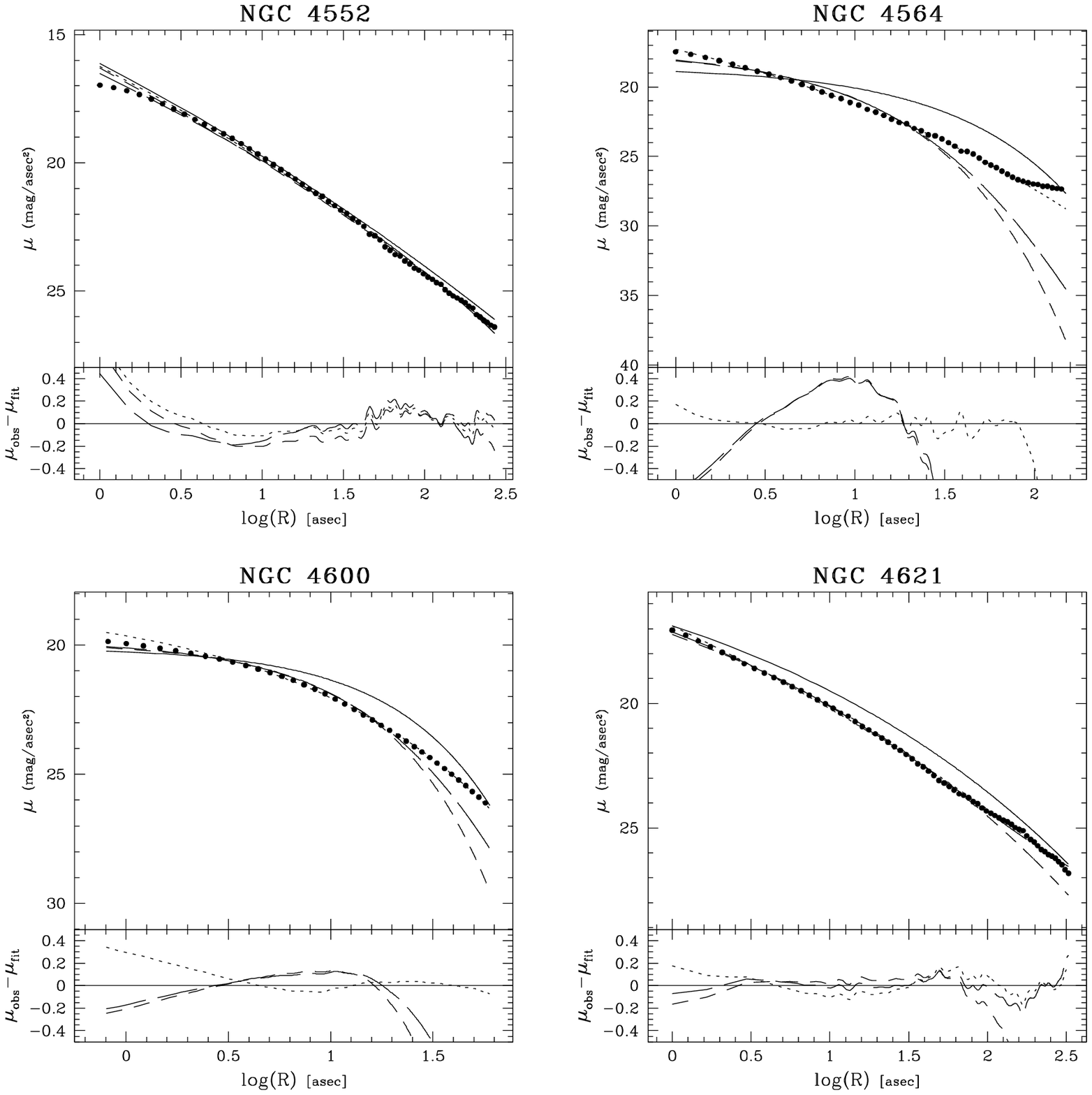}
    \caption{Surface brightness profiles. Continued.}
  \end{center}
\end{figure}

\begin{figure}
  \begin{center}
    \label{fig:mugalG}
    \includegraphics[width=0.5\textwidth]{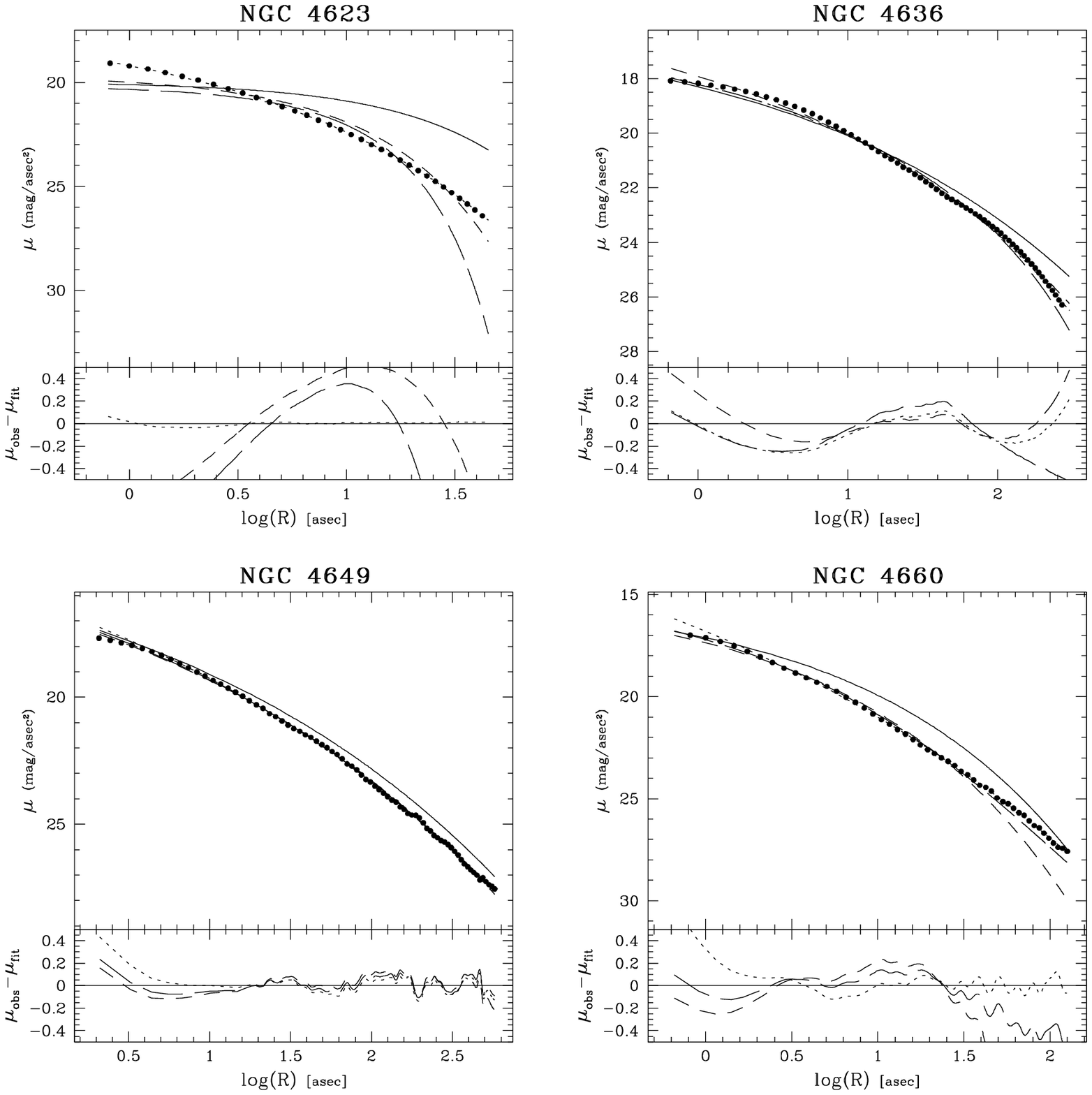}
    \caption{Surface brightness profiles. Continued.}
  \end{center}
\end{figure}

\pagebreak

\section{Lerch $\Phi$ function}
\label{sec:lerch}

The Lerch $\Phi$ function (named after Mathias Lerch, 1860-1922) is defined as
an infinite series (Gradshteyn \& Ryzhik 2000)
\begin{equation}  \label{eq:lerchphi1}
  \Phi(z,a,v) = \sum_{i=0}^{\infty} \frac{z^i}{(v+i)^a}
\end{equation}
where $v+i\neq 0$. In the case studied in equation~(\ref{eq:muphi}) we have
\begin{eqnarray}
  \label{eq:lerchphi2}
  \Phi
  {\textstyle \left\{  1- \frac{\mathcal{F}(a)}{e_0}; 1 ; \frac{1}{nl} \right\} }
  =
  \sum_{i=0}^{\infty}
  \frac{nl}{ 1 + nli}
  \left[ (1+l)
    \Big(1- \frac{e_1}{e_0} \Big)
    \Big(\frac{a}{\aM}\Big)^{l} \right]^i .
\end{eqnarray}
In this case ($a=1$), one of the constraints for $\Phi$ to be finite is that we
must have $|z| = |1 - \mathcal{F}(a)/e_0| < 1$, which corresponds to a
critical radius \ac\ beyond which  $\Phi$ is finite, given by
\begin{equation}
  \label{eq:ac}
  \ac \equiv \frac{\aM}{\left| (1+l)(1-e_0/e_1) \right|^{1/l} }.
\end{equation}
We now may write eq.~(\ref{eq:lerchphi2}) in terms of \ac
\begin{equation}
 \label{eq:lerchphi_ac}
 \Phi
 {\textstyle \left\{  1- \frac{\mathcal{F}(a)}{e_0}; 1 ; \frac{1}{nl} \right\}}
 = \sum_{i=0}^{\infty} \frac{nl}{1+nli}
 \genfrac{(}{)}{}{}{a}{\ac}^{l+i}
\end{equation}
The other constraint is that $1+inl\neq 0$ in equation (\ref{eq:lerchphi_ac})
above, thus $nl\neq \ldots,-2, -1, 0$. When fitting the galaxy eccentricity
profiles to equation (\ref{eq:epsilon}) we must take these constraints into 
account.

\begin{figure}
  \begin{center}
    \includegraphics[width=0.45\textwidth]{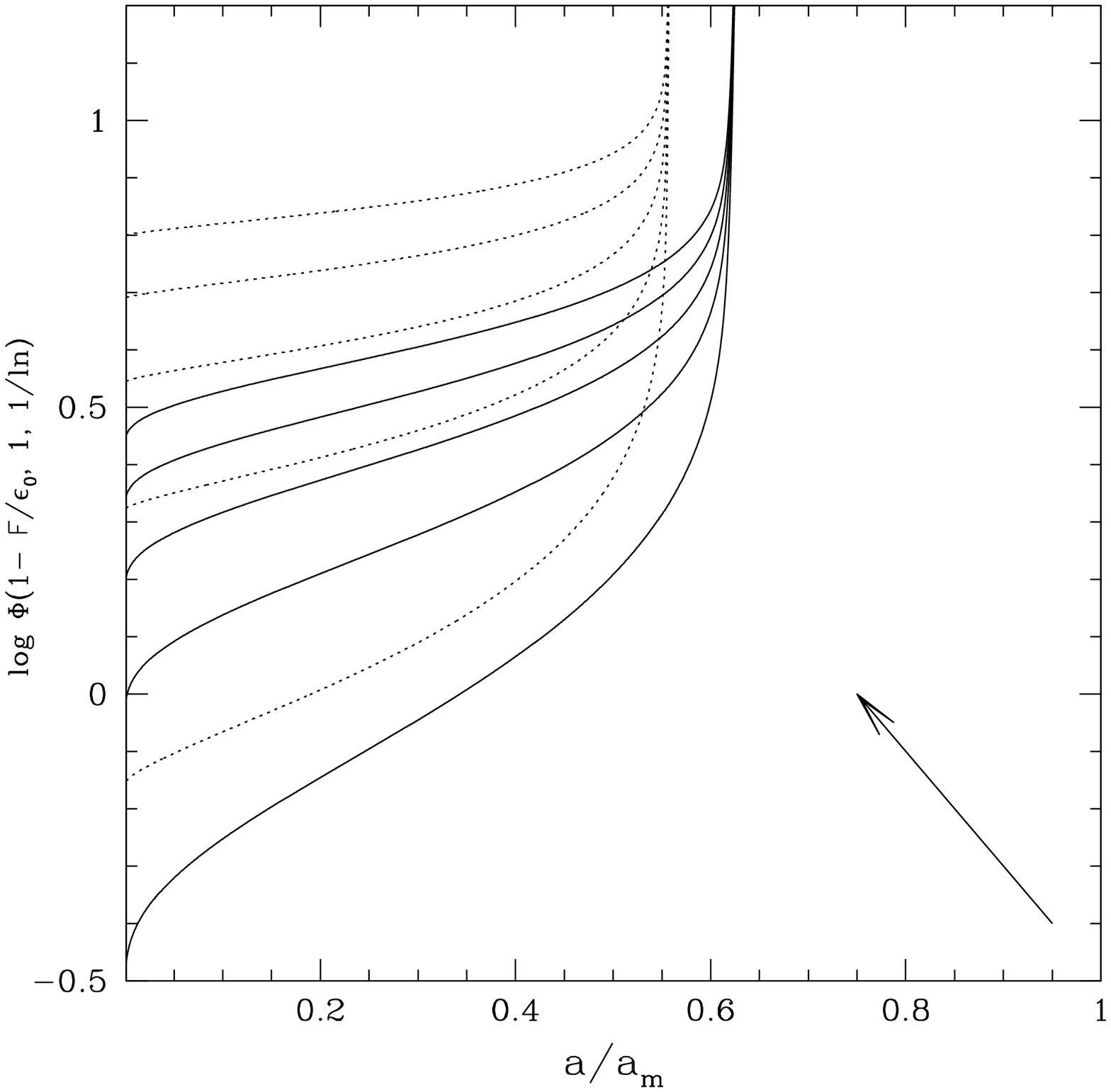}
     \caption{The dependence of the Lerch $\Phi$ function on the \Sersic\
      index $n$. The plotted values are $n=1,\: 3,\: 5,\: 7,\: 9$ increasing
      as indicated by the arrow. The solid lines are for $l=0.3$ and the dotted
      lines for $l=0.7$. In both cases $e_0=0.9$ and $e_1=0.1$. }
    \label{fig:phi-n}
  \end{center}
\end{figure}

\begin{figure}
  \begin{center}
    \includegraphics[width=0.45\textwidth]{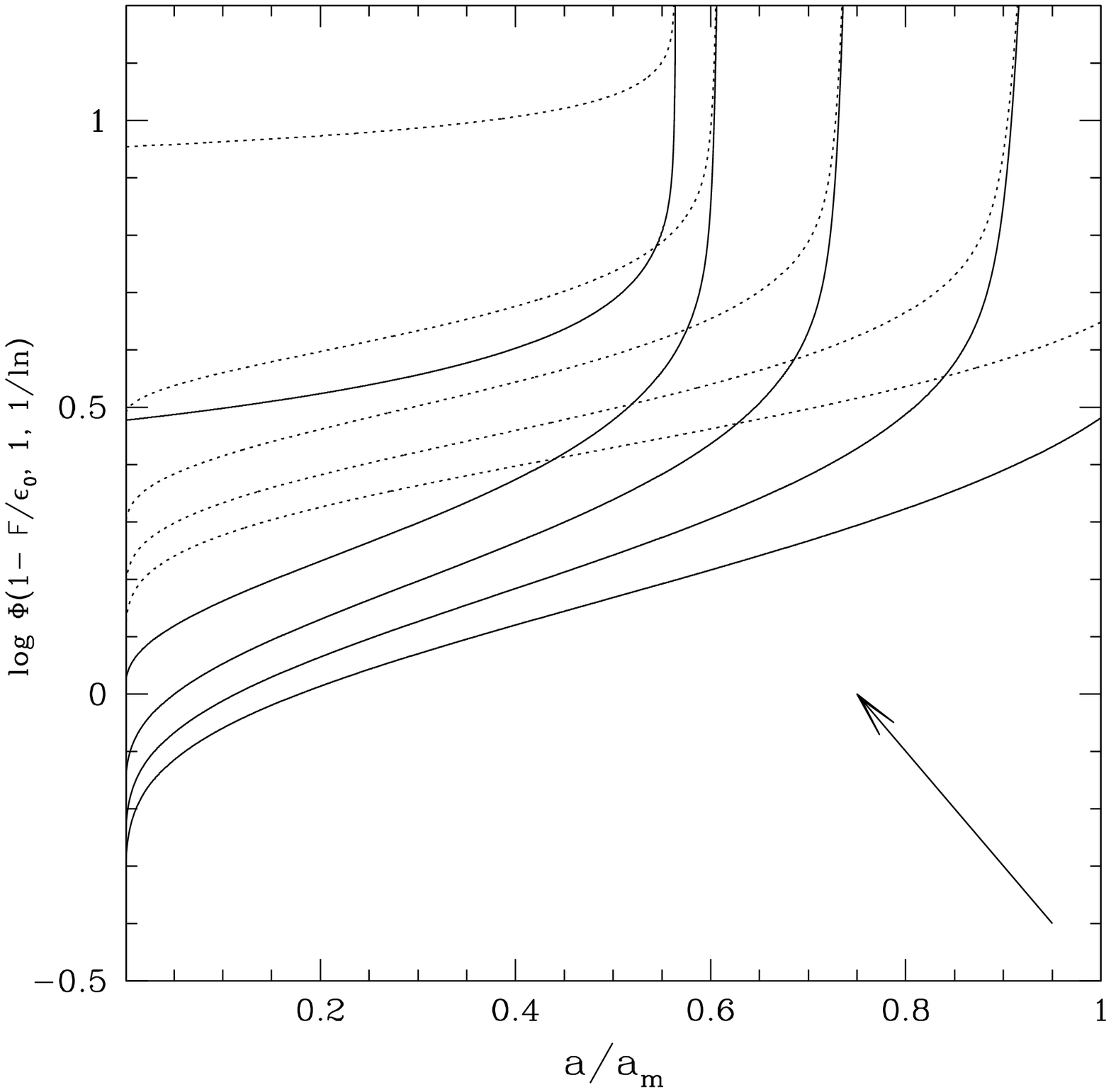}
    \caption{The dependence of the Lerch $\Phi$ function on the eccentricity
      parameter $l$. The plotted values are 
      $l=1,\: 1/3,\: 1/5,\: 1/7,\: 1/9$ increasing as indicated by the arrow. 
      The solid lines are for $n=3$ and the dotted lines for $n=9$. In both 
      cases $e_0=0.9$ and $e_1=0.1$.}
    \label{fig:phi-l}
  \end{center}
\end{figure}

The dependence of the Lerch $\Phi$ function on the $n$ and $l$ parameters is
shown in Figures \ref{fig:phi-n} and \ref{fig:phi-l}.  Figure~\ref{fig:phi-n}
shows how $\Phi_L$ changes for values of $n=1,3,5,7,9$, $n$ raising in the
direction indicated by the arrow.  The solid curves have $l=0.3$ and the dotted
curves have $l=0.7$. The same is true for Figure~\ref{fig:phi-l}, for which we
plot the values $l=1,\; 1/3,\; 1/5,\; 1/7,\; 1/9$, the solid curves having $n=3$
and the dotted curves having $n=9$.  For all cases, $e_0=0.9$ and $e_1=0.1$. The
critical radius \ac\ beyond which the function diverges should be noted. For
example, in Figure \ref{fig:phi-n} the solid line has $\ac/\aM = 0.62$ and the
dotted lines $\ac/\aM = 0.55$, cf.  equation~(\ref{eq:lerchphi_ac}) and since
\ac\ does not depend on $n$ all the curves in Figure~\ref{fig:phi-n} have the
same critical radius.

\begin{thebibliography}{}



\bibitem{bi1}
  Binney, J., Merrifield, M. 1998, Galactic Astronomy.  Princeton:
  Princeton University Press

\bibitem{ca0}
  Caon, N., Capaccioli, M., Rampazzo, R. \ 1990, A\&AS, 86, 429

\bibitem{ca1}
  Caon, N., Capaccioli, M., D'Onofrio, M.\ 1993, MNRAS, 265, 1013 \ (CCD93)

\bibitem{ca2}
  Caon, N., Capaccioli, M., D'Onofrio, M.\ 1994, A\&AS, 106, 199 



\bibitem{ci1}
Ciotti, L. 1991, A\&A, 249, 99

\bibitem{ci2}
  Ciotti, L., Lanzoni, B., Renzini, A. 1996, MNRAS, 282, 1

\bibitem{ci3}
  Ciotti, L., Lanzoni, B.  1997, A\&A, 321, 724

\bibitem{v1}
  de Vaucouleurs, G. 1948, Ann. Astrophys., 11, 247





\bibitem{Grad} 
  Gradshteyn, I. S. and Ryzhik, I. M. Tables of Integrals, Series, and 
  Products, 6th ed. San Diego, CA: Academic Press, 2000


\bibitem{gr1}
  Graham, A.,  Colless, M. 1997, MNRAS, 287, 221

\bibitem{gr2001}
  Graham, A.,  Erwin, P., Caon, N., Trujillo, I. 2001, ApJ, 563, L11

\bibitem{gr2001a}
  Graham, A.,  Trujillo, I. Caon, N., 2001, AJ, 122, 1717





\bibitem{se}
  S\'ersic, J.~L.\ 1968, Cordoba, Argentina: Observatorio Astronomico, 1968

\bibitem{scho}
Schombert, J. M. 1986, ApJS, 60, 603
\end{thebibliography}
\end{document}